\documentclass[runningheads]{llncs}
\usepackage[T1]{fontenc}
\usepackage{graphicx}
\usepackage{booktabs}
\usepackage[misc]{ifsym}
\usepackage{multirow}

\usepackage{mwe}
\usepackage{amsmath}
\begin{document}

\title{LiteSemRAG: Lightweight LLM-Free Semantic-Aware Graph Retrieval for Robust RAG}

\author{Xiao Yue\inst{1} \and
Guangzhi Qu\inst{1} \and
Lige Gan\inst{1}}

\authorrunning{X.Yue et al.}

\institute{Oakland University, Rochester Hills, MI 48309, USA \email{\{xiaoyue,gqu,lgan\}@oakland.edu}
}

\maketitle              

\begin{abstract}

Graph-based Retrieval-Augmented Generation (RAG) has shown great potential for improving multi-level reasoning and structured evidence aggregation. However, existing graph-based RAG frameworks heavily rely on exploiting large language models (LLMs) during indexing and querying, leading to high token consumption, computational cost and latency overhead. In this paper, we propose \textbf{LiteSemRAG}, a lightweight, fully LLM-free, semantic-aware graph retrieval framework. LiteSemRAG constructs a heterogeneous semantic graph by exploiting contextual token-level embeddings, explicitly separating surface lexical representations from context-dependent semantic meanings. To robustly model polysemy, we introduce a dynamic semantic node construction mechanism with chunk-level context aggregation and adaptive anomaly handling. At query stage, LiteSemRAG performs a two-step semantic-aware retrieval process that integrates co-occurrence graph weighting with an isolated semantic recovery mechanism, enabling balanced structural reasoning and semantic coverage. We evaluate LiteSemRAG on three benchmark datasets and experimental results show that LiteSemRAG achieves the best mean reciprocal rank (MRR@10) across all datasets and competitive or superior recall rate (Recall@10) compared to state-of-the-art LLM-based graph RAG systems. Meanwhile, LiteSemRAG consumes zero LLM tokens and achieves substantial efficiency improvements in both indexing and querying due to the elimination of LLM usage. These results demonstrate the effectiveness of LiteSemRAG, indicating that a strong semantic-aware graph retrieval framework can be achieved without relying on LLM-based approaches.

\end{abstract}

\section{Introduction}
Retrieval-Augmented Generation (RAG) \cite{lewis2020retrieval,borgeaud2022improving,izacard2021leveraging,zhao2026retrieval} has become a fundamental paradigm for improving factual grounding and knowledge integration in large language models (LLMs) \cite{brown2020language,devlin2019bert}.  By augmenting generation with retrieval over external corpora, RAG systems can improve factuality, adaptability, and domain generalization in knowledge-intensive tasks \cite{guu2020retrieval}. Therefore, the quality and efficiency of the retrieval component have become critical factors in RAG systems. Early RAG systems mainly rely on flat chunk-level retrieval using dense or sparse representations~\cite{karpukhin2020dense,khattab2020colbert}. However, RAG systems with flat chunk-level retrieval often struggle with document structure, semantic variation, and multi-hop evidence dependencies. To address these limitations, recent work has introduced structure-aware and graph-based retrieval frameworks \cite{sarthi2024raptor,gutierrez2024hipporag,edge2024local,guo2024lightrag}. They construct hierarchical or graph-based index structures to enable multi-level reasoning and structured evidence aggregation. These approaches demonstrate that structured indexing can substantially improve retrieval effectiveness, particularly in complex reasoning scenarios. However, many existing graph-based RAG frameworks heavily rely on LLMs during both indexing and querying. Although LLMs provide powerful capabilities for semantic extraction, summarization, and reasoning, utilization of LLMs also introduces substantial computational overhead, high token consumption, latency variability, and sensitivity to the specific underlying model. These factors can significantly limit the practicality of such systems in resource-constrained or large-scale scenarios. NoLLMRAG~\cite{yunollmrag} first attempts to eliminate LLM dependence in graph-based RAG. However, it only relies on surface-level lexical and static statistical structures, which limits contextual and semantic understanding.

In this paper, we propose \textbf{LiteSemRAG}, a lightweight, LLM-free, semantic-aware graph retrieval framework. Unlike existing graph-based RAG frameworks that depend on LLMs for semantic extraction and reasoning, LiteSemRAG demonstrates that strong semantic-aware retrieval can be achieved using purely embedding-based semantic graph construction, significantly reducing cost and improving scalability. LiteSemRAG constructs a heterogeneous semantic graph directly by exploiting contextual token-level embeddings. The heterogeneous semantic graph explicitly separates surface lexical representations (token nodes) from context-dependent semantic meanings (semantic nodes), enabling dynamic modeling of polysemy without relying on LLMs. To enhance robustness, LiteSemRAG introduces chunk-level context aggregation and adaptive anomaly handling to stabilize semantic node generation under weak contextual information and incremental indexing. At query stage, LiteSemRAG performs a two-step semantic-aware retrieval process. First, it builds a query-specific semantic co-occurrence graph that integrates lexical constraints, semantic similarity, and structural co-occurrence weighting for chunk ranking. Then it introduces an isolated semantic recovery mechanism that propagates structural semantic weight within token families to recover semantically relevant but structurally unsupported evidence. This design allows LiteSemRAG to preserve both structural reasoning and semantic coverage while maintaining computational efficiency.
We evaluate LiteSemRAG on three widely used benchmarks: HotpotQA, SciFact, and FEVER. Experimental results show that LiteSemRAG achieves the best MRR@10 across all
datasets and competitive or superior Recall@10 to two state-of-the-art LLM-based graph RAG systems. This design makes LiteSemRAG particularly suitable for large-scale retrieval systems where LLM-based indexing introduces prohibitive computational and monetary costs.
The contributions of this paper are summarized as follows:
\begin{itemize}
    \item We propose the first LLM-free semantic-aware graph retrieval framework that explicitly models token polysemy through contextual embeddings.
    \item We introduce dynamic semantic node construction with chunk-level context aggregation and adaptive anomaly handling mechanisms to robustly derive semantic nodes from contextual token embeddings.
    \item We design a two-step semantic-aware retrieval strategy combining co-occurrence graph semantic retrieval and isolated semantic recovery mechanism.
    \item We demonstrate that LiteSemRAG achieves competitive or superior retrieval performance compared to LLM-based graph RAG systems while significantly improving efficiency without any LLM token consumption.
\end{itemize}

\section{Related Work}

\subsection{Retrieval-Augmented Generation}

Retrieval-augmented generation (RAG) \cite{lewis2020retrieval} combines a parametric generator with a non-parametric retriever on an external corpus. RAG has become a core component for improving performance in knowledge-intensive natural language processing tasks. However, many RAG frameworks still rely on flat chunk retrieval, which can struggle to capture document structure, semantic variation, and multi-hop evidence dependencies \cite{karpukhin2020dense}. Some works are proposed to improve the retrieval component through dense or sparse neural representations. Karpukhin et al. \cite{karpukhin2020dense} proposed a dual-encoder dense retriever that maps questions and passages into a shared embedding space for nearest-neighbor search. ColBERT \cite{khattab2020colbert} is proposed by exploiting late interaction to preserve token-level matching signals while remaining efficient for large-scale retrieval. These methods substantially improve retrieval quality, but they generally retrieve over flat format collections and do not explicitly induce graph-structured semantic units.


\subsection{Structure-Aware and Graph-Based RAG}
Recent work has explored structured retrieval to better support long-context reasoning and multi-hop evidence aggregation. RAPTOR \cite{sarthi2024raptor} builds a hierarchical tree of clustered and summarized text units, enabling retrieval at multiple abstraction levels. GraphRAG \cite{edge2024local} constructs an entity-centric graph and uses LLM-based extraction and community summaries to support global and local query answering. HippoRAG \cite{gutierrez2024hipporag} combines knowledge graphs with personalized PageRank to improve multi-hop retrieval efficiency. LightRAG \cite{guo2024lightrag} further simplifies graph-based RAG with dual-level retrieval over low-level and high-level knowledge structures. These methods show that structured indexing can improve retrieval quality beyond flat chunk search, especially for complex reasoning tasks. However, the dominant graph-based RAG frameworks still depend on LLMs during indexing and querying, which increases latency, cost, and reproducibility sensitivity to the underlying model. NoLLMRAG \cite{yunollmrag} is the first graph-based RAG framework that eliminates reliance on LLMs during indexing and querying. However, due to absence of explicit semantic modeling, it only relies on surface-level lexical and
static statistical structures, which can significantly limit its effectiveness in complex or multi-hop scenarios.

\section{Approach}
\begin{figure}[htb]
    \centering
    \includegraphics[width=1\linewidth]{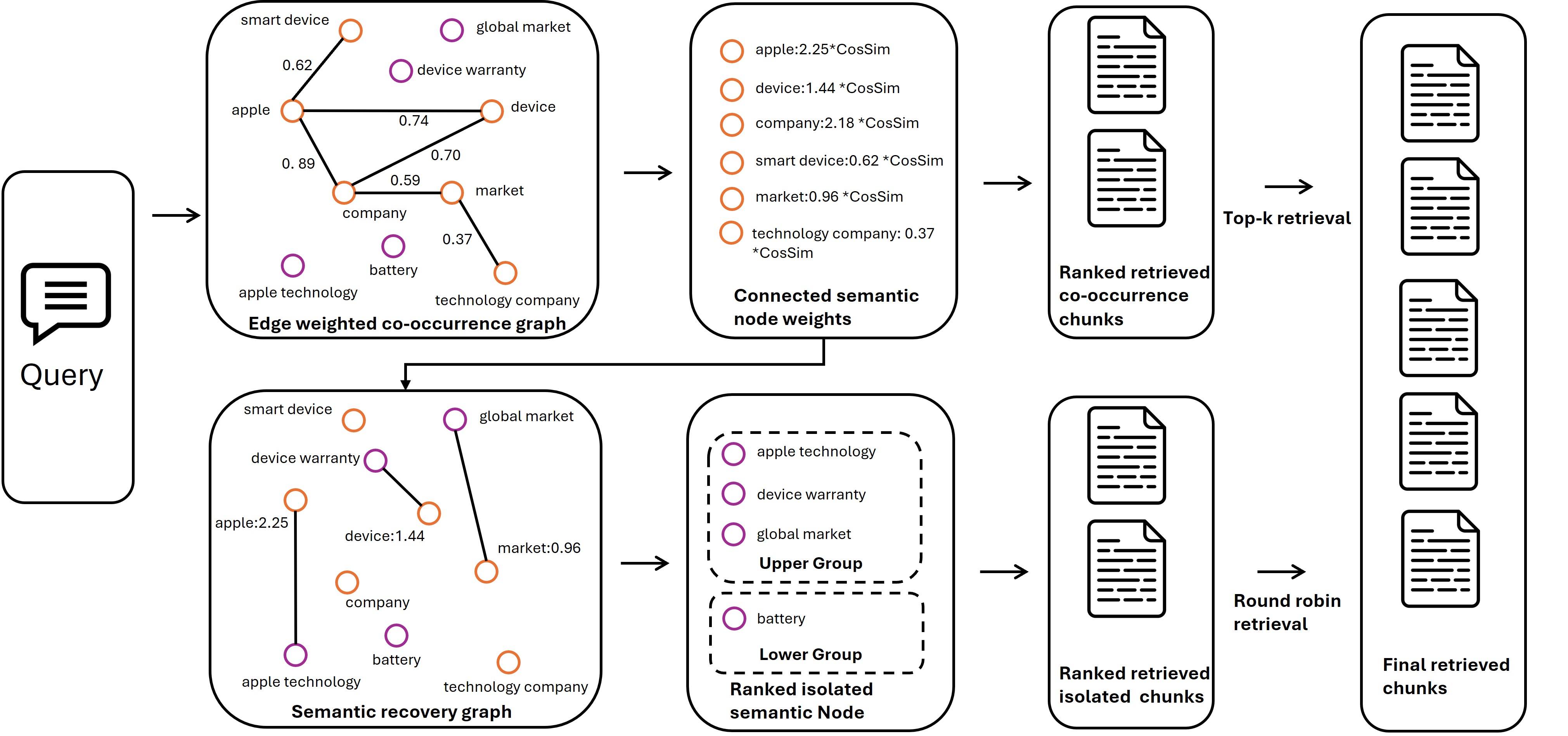}
    \caption{Overview of the two-step semantic-aware retrieval process}
    \label{query}
\end{figure}

LiteSemRAG is a fully LLM-free semantic retrieval framework that constructs a heterogeneous semantic graph to preserve contextual semantic meaning while maintaining efficient retrieval. The framework consists of two main components: \textbf{Heterogeneous Semantic Graph Index}, where semantic-aware structures are derived from contextual token-level embeddings, and \textbf{Two-step Semantic-aware Retrieval}, where retrieval is performed through a two-step co-occurrence graph-based semantic reasoning process. Figure~\ref{query} illustrates the workflow of the two-step semantic-aware retrieval. It consists of two stages: \emph{Co-occurrence Graph Semantic Retrieval}, which captures semantic relationships through a semantic co-occurrence graph, and \emph{Isolated Semantic Recovery}, which recovers semantically relevant information that may be missed in the first stage. The ranked chunks retrieved from these two stages are then combined to produce the final results. Chunks obtained from Co-occurrence Graph Semantic Retrieval are prioritized over those from Isolated Semantic Recovery in the final ranking, while a user-defined hyperparameter controls the proportion of chunks selected from each stage.
\subsection{Graph Structure}
\begin{figure}[htb]
    \centering
    \includegraphics[width=0.8\linewidth]{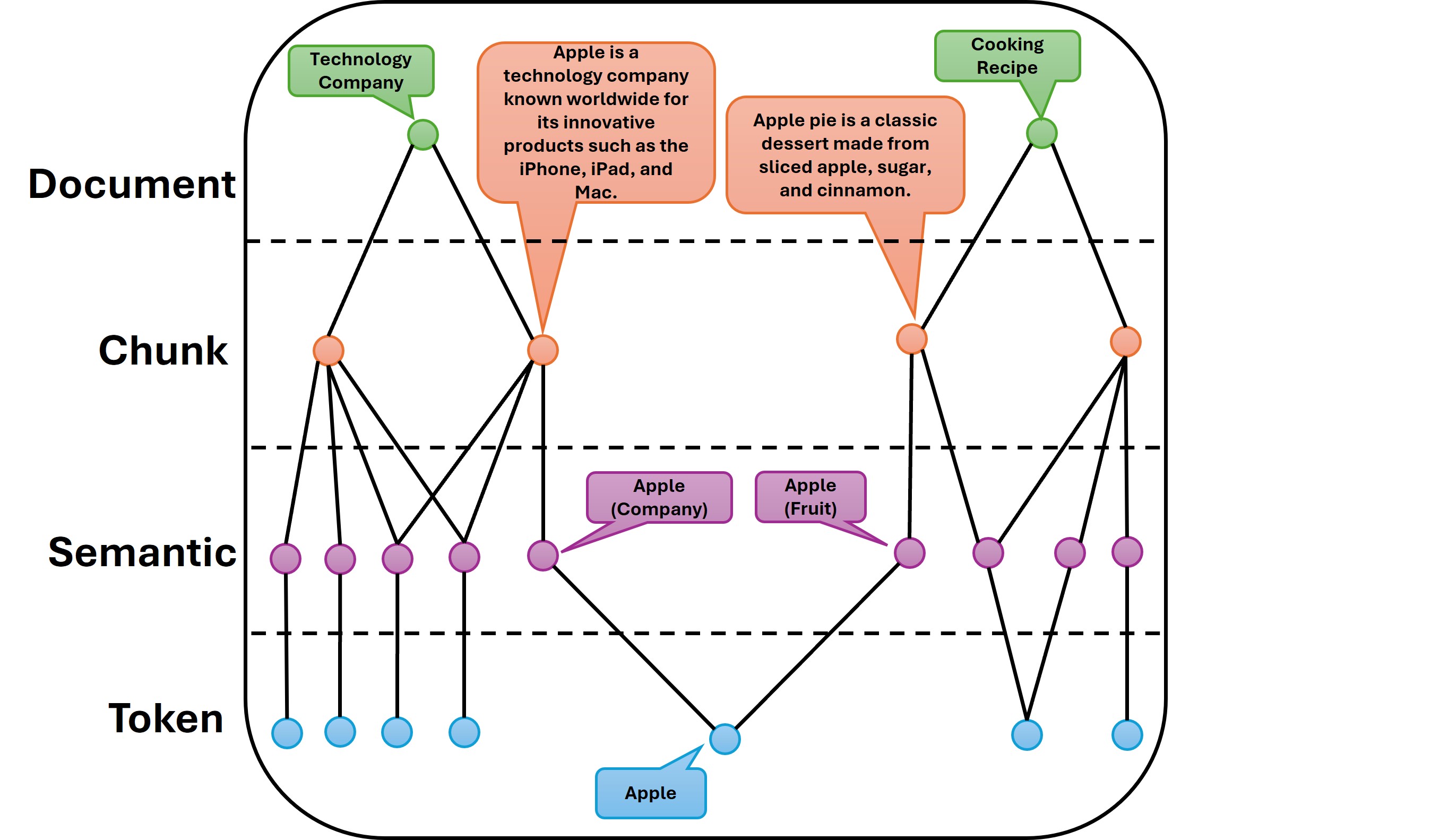}
    \caption{Example of the multi-layer heterogeneous semantic graph index}
    \label{graph}
\end{figure}
We define a multi-layer heterogeneous semantic graph as $G = (V, E)$, where the node set is defined as $V = D \cup C \cup S \cup T$. The node types are defined as:
\begin{itemize}
    \item \textbf{Document nodes} ($D$): each represents a document in the corpus.
    \item \textbf{Chunk nodes} ($C$): each represents a text chunk derived from a document.
    \item \textbf{Semantic nodes} ($S$): each represents a specific semantic meaning of a token node.
    \item \textbf{Token nodes} ($T$): each represents a unique surface textual element (entity, phrase, or token).
\end{itemize}
The edge set $E$ consists of bidirectional edges of three types:
\begin{itemize}
    \item Document--Chunk edges: indicating that a chunk belongs to a document.
    \item Chunk--Semantic edges: indicating that a semantic meaning appears in a chunk.
    \item Semantic--Token edges: linking each semantic node to its corresponding surface token.
\end{itemize}

\subsection{Heterogeneous Semantic Graph Index}
Given a document collection, LiteSemRAG first splits each document into chunks. From each chunk, important textual elements (entities, phrases, and tokens) are extracted by NLP tools. These elements are filtered and encoded using a token-level text encoder to obtain contextual semantic embeddings. Semantic nodes are dynamically constructed to represent distinct contextual meanings based on contextual semantic embeddings. The final index graph is organized as a multi-layer heterogeneous semantic graph connecting documents, chunks, semantic nodes, and token nodes. 
\subsubsection{Dynamic semantic node creation}

A key component of LiteSemRAG is the dynamic construction of semantic nodes from contextual semantic embeddings. Instead of assigning a fixed representation to each surface token element, LiteSemRAG creates distinct semantic nodes that capture context-dependent meanings of the same token element. This enables explicit modeling of polysemy without using LLM-based parsing. For each extracted token $t$, we collect its contextual semantic embeddings across all chunk occurrences. Let
$E_t = \{ e_1, e_2, \dots, e_n \}$ denote the set of normalized contextual semantic embeddings for a token node $t$. To determine whether a token exhibits multiple semantic meanings, we measure the concentration of its embedding distribution by computing the dispersion score (S-mean) which is defined as 
\[
S\text{-mean}(t) = \frac{1}{n} \sum_{i=1}^{n} \cos(e_i, \bar{e}_t).
\]
where $\bar{e}_t = \frac{1}{n} \sum_{i=1}^{n} e_i$. A high S-mean value indicates that embeddings are tightly concentrated around a single semantic direction, while a low S-mean suggests multiple contextual clusters. To avoid unnecessary splitting of highly frequent tokens, we jointly consider the inverse document frequency (IDF). We conduct clustering-based semantic induction on a token $t$ only if
$
IDF(t) > \tau_{idf} \quad \text{and} \quad S\text{-mean}(t) < \tau_{disp}
$,
where $\tau_{\text{idf}}$ and $\tau_{\text{disp}}$ are predefined thresholds.

\subsubsection{Clustering-Based Semantic Induction.}
If a token is identified as possible to be multi-semantic, we apply density-based clustering algorithms, such as HDBSCAN, on $E_t$ to determine the number of semantic clusters. Each cluster corresponds to one semantic node. For cluster $C_k \subset E_t$, its semantic anchor embedding is defined as:
\[
e_{s_k} = \frac{1}{|C_k|} \sum_{e_i \in C_k} e_i.
\]
If the token is determined as single-semantic by either the threshold criteria or clustering algorithm, only one semantic node is created. The semantic anchor embedding for this node is computed by averaging all contextual embeddings associated with the token. After semantic nodes are formed, chunk nodes previously connected to this token node are re-linked to the corresponding semantic nodes. The token node remains connected to all of its semantic nodes, serving as a lexical anchor. Once the semantic nodes are generated, raw contextual semantic embeddings assigned to semantic nodes are discarded to save memory space while the averaged semantic representations are retained to ensure that semantic information is preserved.

\subsubsection{Chunk-Level Context Aggregation for Robust Semantic Discovery}
\begin{figure}
    \centering
    \includegraphics[width=1\linewidth]{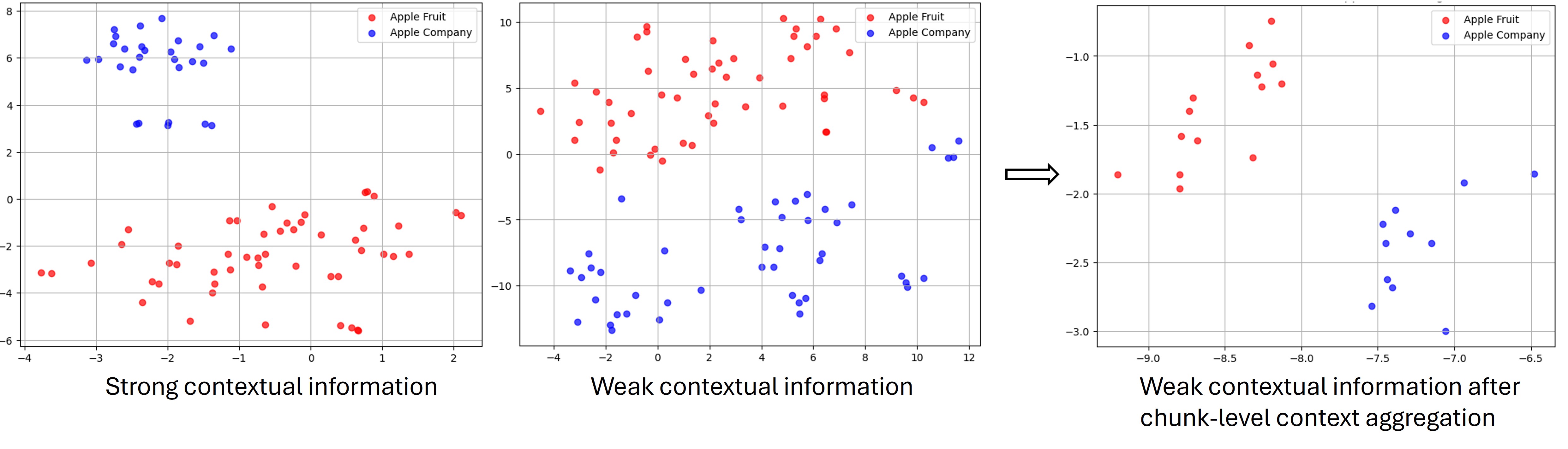}
    \caption{Embedding distribution of strong contextual information (left), weak contextual information (middle), and after aggregation (right)}
    \label{embed}
\end{figure}

In practice, we observe that contextual semantic embeddings of polysemous tokens may not form clearly separable clusters in the embedding space, especially when local contextual information is weak. For example, the token \emph{apple} may appear in both technological and fruit-related contexts. Although a separating boundary exists in principle, unsupervised clustering may fail to identify it when individual token embeddings contain limited contextual information, as shown in Figure \ref{embed}.

To mitigate this issue, we introduce a chunk-level context aggregation strategy prior to clustering. Specifically, for token $t$, suppose its contextual semantic embeddings are $E_t = \{ e_1, e_2, \dots, e_n \}$, where each embedding $e_i$ is associated with a particular chunk. If the initial clustering attempt does not produce stable or meaningful clusters, we perform aggregation at the chunk level. Let $\mathcal{C}_t$ denote the set of chunks containing token $t$. For each chunk $c \in \mathcal{C}_t$, we compute a chunk-level embedding:
\[
\tilde{e}_c = \frac{1}{|E_{t,c}|} \sum_{e_i \in E_{t,c}} e_i,
\]
where $E_{t,c}$ denotes the embeddings of token $t$ in chunk $c$. These aggregated embeddings replace the original embeddings for clustering.  Averaging embeddings within same chunk reduces intra-chunk variance while preserving inter-chunk semantic differences. This aggregation enhances cluster separability by amplifying consistent contextual signals, as shown in Figure \ref{embed}. This strategy improves the robustness of dynamic semantic node creation, particularly for tokens whose individual contextual information is weak.

\subsubsection{Adaptive Anomaly Handling}
To support evolving semantic distributions during incremental indexing, LiteSemRAG introduces an anomaly handling mechanism. When a new contextual semantic embedding $e_{\text{new}}$ is derived for token $t$ who already has any semantic nodes, it is compared against anchor embeddings of existing semantic nodes. The embedding is temporarily stored in an anomaly set if
$
\max_{s \in S(t)} \cos(e_{\text{new}}, e_s) < \tau_{\text{anomaly}}
$.
Instead of using a fixed similarity threshold to detect emerging meanings, we adopt a percentile-based adaptive strategy. We define the adaptive anomaly threshold $\tau_{\text{anomaly}}$ of token $t$ as the \emph{n}-th percentile of $\mathcal{S}_t$, where $n$ is a user-selected hyperparameter. The \emph{n}-th percentile refers to the similarity value below which \emph{n\%} of the elements in $\mathcal{S}_t$ fall. In other words, it represents a threshold such that only a small portion of the lowest-similarity embeddings are considered anomalous. This percentile-based strategy allows the threshold to adapt to the intrinsic dispersion of each token's semantic distribution, avoiding rigid global cutoffs. Additionally, certain clustering algorithms may generate outlier points that do not belong to any cluster. These outliers are also considered anomalies and added to the anomaly set. When the anomaly set size exceeds a preset threshold, reclustering is performed. If a new cluster is identified, a new corresponding semantic node is created; otherwise, all anomaly embeddings are assigned to their nearest semantic nodes.

\subsubsection{Design Rationale}
This dynamic semantic node creation process allows LiteSemRAG to: 1)Explicitly model polysemy without using LLM, 2) Adapt to semantic drift during incremental indexing, 3) Maintain computational and memory efficiency. LiteSemRAG separates surface lexical representation from contextual semantic representation. Token nodes act as lexical anchor points, while semantic nodes capture contextual meanings derived from contextual semantic embeddings. 

\subsection{Co-occurrence Graph Semantic Retrieval}
At query stage, we first extract token elements from the query text using the same preprocessing pipeline in the index stage. Each extracted token element $t_q$ is encoded into a contextual semantic embedding $e_q$ using the same token-level text encoder. 

\subsubsection{Multi-Level Semantic Matching}
For each query token $t_q$ with embedding $e_q$, LiteSemRAG directly retrieves candidate \textbf{semantic nodes}. Matching is performed at three levels, defined by the textual properties of the corresponding token nodes:

\begin{itemize}
    \item \textbf{Exact match:} Semantic nodes whose associated token node is identical to $t_q$.
    
    \item \textbf{Partial match:} Semantic nodes whose associated token node contains $t_q$ as a substring.
    
    \item \textbf{Similarity match:} Semantic nodes whose embeddings are similar to $e_q$ according to cosine similarity.
\end{itemize}
Each matching level is assigned a predefined node-level weight:
$
\alpha_{\text{exact}} > \alpha_{\text{partial}} > \alpha_{\text{similarity}}.
$
For exact and partial matching, candidate semantic nodes are first identified through their linked token nodes. Specifically, exact matching retrieves the token node whose surface form is identical to the query token $t_q$, while partial matching retrieves token nodes whose surface forms contain $t_q$ as a substring. For each matched token node $t$, only its best semantic node is retained based on embedding similarity:
$
s^*(t) = \arg\max_{s \in S(t)} \cos(e_q, e_s),
$
where $\mathcal{S}(t)$ denotes the set of semantic nodes linked to token node $t$, $e_q$ is the embedding of the query token, and $e_s$ is the anchor embedding of semantic node $s$.
Since exact matching corresponds to a unique token node, it yields a single semantic node. In contrast, partial matching may retrieve multiple token nodes, each contributing its best semantic node. To control the number of candidates, we retain at most the top-$k$ semantic nodes ranked by similarity, where $k$ is a user-specified hyperparameter.
For similarity matching, the search is performed directly over semantic nodes without requiring token-level constraints. Semantic nodes are also ranked according to cosine similarity, and the top-$k$ most similar semantic nodes are retained. This design allows retrieval to operate directly at the semantic level while preserving lexical precision through token-level constraints.

\subsubsection{Co-occurrence graph node weighting}
To capture relationships among matched semantic nodes, we construct a query-specific semantic co-occurrence graph. For semantic nodes $i$ and $j$, the edge weight is defined as:
\[
w_{ij} =
\frac{|C(i) \cap C(j)|}
{\sqrt{|C(i)| \cdot |C(j)|}},
\]
where $C(i)$ denotes the set of chunks containing semantic node $i$. This cosine-style normalized co-occurrence metric suppresses globally frequent semantics while emphasizing meaningful associations.
For each semantic node $s$, we compute its node weight:
\[
W(s) =
\left(
\sum_{j \in N(s)} w_{sj}
\right)
\cdot
\alpha_{\text{level}(s)}
\cdot
\cos(e_q, e_s),
\]
where $N(s)$ denotes neighboring semantic nodes in the co-occurrence graph, and $\alpha_{\text{level}(s)}$ is the node-level weight corresponding to its semantic matching level. This formulation integrates co-occurrence graph structural information, lexical matching confidence, and semantic similarity.

\subsubsection{Chunk Scoring}
\label{chunk scoring}
Chunks connected to the selected semantic nodes are ranked by using a BM25-style semantic-level score:
\[
\text{Score}(c) =
\sum_{s \in S(c)}
W(s)
\cdot
G(s)
\cdot
\frac{f(s,c) (k_1 + 1)}
{f(s,c) + k_1 \left(1 - b + b \frac{|c|}{\text{avgcl}}\right)},
\]
In the above formulation, $S(c)$ denotes the set of semantic nodes that appear in chunk $c$. The term $W(s)$ represents the weight of semantic node $s$. The quantity $G(s)$ denotes the chunk-level discriminative weight of semantic node $s$, computed based on its distribution across chunks (such as Inverse Document Frequency). The term $f(s,c)$ denotes the frequency of semantic node $s$ in chunk $c$. The symbol $|c|$ represents the length of chunk $c$, and $\text{avgcl}$ denotes the average chunk length across the corpus. The hyperparameters $k_1$ and $b$ follow the standard BM25 formulation and control term-frequency saturation and length normalization, respectively. Note that each semantic node contributes at most once to the semantic-level score, even if it appears multiple times within the same chunk.

\subsection{Isolated Semantic Recovery}
While co-occurrence graph semantic retrieval captures structurally supported semantic relationships, certain matched semantic nodes may remain isolated from the co-occurrence graph. This occurs when a semantic node matches the query but does not participate in any significant co-occurrence relationship with other matched semantic nodes. To avoid missing potential relevant evidence, we introduce an \emph{Isolated Semantic Recovery} stage. Note that each isolated semantic node $s \in S_q^{(\text{iso})}$ corresponds to exactly one query-matched semantic meaning but does not share co-occurrence edges with other semantic nodes.

\subsubsection{Token-Family Weight Propagation}
Token elements serve as lexical anchors at the multi-level semantic matching phase. As a result, multiple semantic nodes in the co-occurrence graph may correspond to the same original token element. For an isolated semantic node $s$, if there exist semantic nodes 
$\{s'_i\}$ in the co-occurrence graph that share the same original token 
as $s$, we propagate their node weights to $s$. Specifically, the propagated weight assigned to $s$ is defined as the sum of the co-occurrence graph node weights of all such matching nodes:
\[
W_{\text{prop}}(s) = \sum_{s'_i \in \mathcal{M}(s)} W(s'_i),
\]
where $\mathcal{M}(s)$ denotes the set of semantic nodes in the co-occurrence  graph that share the same original token as $s$. If $\mathcal{M}(s)$ is empty, no weight is propagated and $s$ does not 
receive any co-occurrence node weight.
\subsubsection{Isolated Node Ranking}
We divide isolated semantic nodes into two groups: \textbf{Upper Group}: Semantic nodes receiving propagated co-occurrence weight;    \textbf{Lower Group}: Semantic nodes without propagated co-occurrence weight.
Nodes in Upper Group are ranked by $\text{Score}(s) = W_{\text{prop}}(s) \cdot \cos(e_q, e_s)$, where $e_q$ is the embedding of query token and $e_s$ is the anchor embedding of semantic node, while nodes in Lower Group are ranked solely by semantic similarity $\text{Score} = \cos(e_q, e_s)$. Two groups are combined while Upper Group is ranked ahead of Lower group, reflecting the assumption that nodes inheriting structural importance are more likely to be relevant.
\subsubsection{Chunk Ranking and Retrieval} 
For each ranked isolated semantic node $s$ in the combined group, we collect its associated chunks and rank them independently using the same BM25-style semantic-level scoring function in section \ref{chunk scoring}. To promote semantic diversity in the final retrieval results, we employ a round-robin selection strategy across the ranked semantic nodes. Specifically, we iteratively traverse the ordered list of semantic nodes and select the next K highest-ranked chunk from each node in turn. After reaching the last node, the process cycles back to the first node and continues in the same manner. This cyclic retrieval procedure repeats until the desired number of chunks is obtained or all candidate chunks have been exhausted.

Isolated Semantic Recovery serves as a structured fallback mechanism. It leverages token-level relationships to propagate semantic importance in co-occurrence graph within token families. This allows the framework to recover semantically relevant but structurally unsupported evidence, particularly in cases where co-occurrence relationships are sparse or incomplete. 
\subsection{Extensibility of the Index Graph}
In the current framework, we restrict retrieval to co-occurrence-aware semantic ranking and isolated semantic recovery to maintain computational efficiency and robustness. However, the underlying structure of LiteSemRAG’s heterogeneous semantic index graph naturally supports more complex graph-based reasoning operations. The heterogeneous graph structure connects documents, chunks, semantic nodes, and token nodes in a bidirectional manner. This design enables flexible traversal across different elements and levels, allowing more advanced operations, including multi-hop semantic expansion, constrained subgraph extraction, and path-based relevance scoring. Therefore, LiteSemRAG should be viewed not only as a retrieval algorithm but also as a structured semantic indexing framework. The index graph provides a foundation for future extensions that incorporate deeper graph traversal strategies or reasoning mechanisms without requiring LLM-based graph construction.

\section{Experiments}
We evaluate LiteSemRAG against representative LLM-based graph RAG systems in terms of retrieval effectiveness and computational efficiency. Our experiments aim to answer the following questions:
\begin{itemize}
    \item Does LiteSemRAG achieve competitive retrieval performance compared to LLM-based graph RAG systems?
    \item How does LiteSemRAG compare in indexing and querying efficiency?
\end{itemize}
\subsection{Experimental Setup}
\subsubsection{Dataset}
We conduct experiments on three widely used retrieval benchmarks from the BEIR dataset \cite{thakur2021beir}: \textbf{HotpotQA} \cite{yang2018hotpotqa}: A multi-hop question answering dataset requiring reasoning across multiple documents; \textbf{SciFact} \cite{Wadden2020FactOF}: A scientific fact verification dataset focusing on evidence retrieval; \textbf{FEVER}: A large-scale fact verification dataset where claims must be supported or refuted using evidence sentences retrieved from Wikipedia.
For all datasets, documents are split into fixed-length chunks using the same preprocessing strategy across all compared methods to ensure fairness.
\subsubsection{Baseline and Setup}
We compare LiteSemRAG with two state-of-the-art LLM-based graph RAG systems: \textbf{GraphRAG} and \textbf{LightRAG}. Since they require LLMs for indexing and querying, we evaluate them under two LLM settings: \textbf{GPT-4o-mini} (API-based) and \textbf{LLaMA3-8B-Instruct} (locally deployed).
The two LLM-based RAG systems use identical LLM configurations within each setting. Both GraphRAG and LightRAG utilize \textbf{text-embedding-3-small} as the text embedding encoder. While LiteSemRAG uses \textbf{deberta-v3-large} as the token-level text encoder. The encoder difference arises because LiteSemRAG requires token-level contextual embeddings from a token-level text encoder, whereas the compared systems rely on sentence-level embeddings derived from sentence-level encoders. All experiments are conducted on a server equipped with two AMD EPYC 9354 CPUs and one AMD MI210 GPU. 

\subsubsection{Evaluation Scope}
LiteSemRAG is designed as a retrieval framework rather than an end-to-end RAG system. Therefore, we evaluate only the quality of retrieved chunks. For fair comparison, we directly measure retrieval effectiveness using Recall@10 and MRR@10 based on ground-truth relevant chunks. We do not compare final answer quality obtained by feeding retrieved chunks into LLMs, since such evaluation would introduce additional variability from generation models and obscure differences in retrieval performance. This design isolates the retrieval component and ensures that performance differences reflect retrieval capability rather than downstream LLM reasoning or generation quality.
\subsubsection{Evaluation Metrics}
We evaluate retrieval quality by using \emph{Recall@10} and \emph{MRR@10}. For each query $q$ in a query dataset $Q$, let $R_q^{(10)}$ denote the ranked top-10 retrieved chunks, and let $G_q$ denote the set of chunks of gold (ground-truth) documents associated with $q$. In HotpotQA dataset, each query has exactly two gold documents. In SciFact and Fever datasets, each query has between one and five gold documents. We compute Recall@10 using a micro-averaged formulation:
\[
\text{Recall@10} =
\frac{
\sum_{q \in Q} | R_q^{(10)} \cap G_q |
}{
\sum_{q \in Q} | G_q |
}.
\]
This measures the proportion of gold documents that are successfully retrieved within the top 10 results across the entire dataset.

For each query $q$ in a query dataset $Q$, let $\text{rank}_q$ denote the rank position of the first retrieved document that belongs to $G_q$ within the top 10 results. If no gold document appears in the top 10, the reciprocal rank is defined as 0. The reciprocal rank is:
\[
\text{RR@10}(q) =
\begin{cases}
\frac{1}{\text{rank}_q}, & \text{if a gold document appears in top 10}, \\
0, & \text{otherwise}.
\end{cases}
\]The Mean Reciprocal Rank at 10 is computed as:
\[
\text{MRR@10} =
\frac{1}{|Q|}
\sum_{q \in Q}
\text{RR@10}(q).
\]
Recall@10 measures how completely the required evidence documents are retrieved, while MRR@10 evaluates how early at least one relevant document appears in the ranked chunk list. In addition, we report four efficiency metrics: Average Indexing Time per document (AIT), Average Query Time (AQT), average Index Token Consumption per document (ITC), average Query Token Consumption (QTC). These metrics quantify both computational and cost efficiency.

\subsection{Retrieval Performance}
\begin{table}[]
\centering
\caption{Retrieval results on three datasets}
\label{retrieval results}
\begin{tabular}{lcccccc}
\hline
\multicolumn{1}{c}{\multirow{2}{*}{Model}} & \multicolumn{2}{c}{HotpotQA} & \multicolumn{2}{c}{SciFact} & \multicolumn{2}{c}{Fever}                                  \\ \cline{2-7} 
\multicolumn{1}{c}{}                       & Recall@10      & MRR@10      & Recall@10      & MRR@10     & \multicolumn{1}{c}{Recall@10} & \multicolumn{1}{c}{MRR@10} \\ \hline
GraphRAG (GPT)                            & 55.9\%         & N/A         & 87.4\%         & N/A        &          45.8\%                      & \multicolumn{1}{c}{N/A}    \\
GraphRAG (LLaMA)                           & ETL            & N/A         & ETL            & N/A        & ETL                           & \multicolumn{1}{c}{N/A}    \\
LightRAG (GPT)                             & 71.5\%         & 0.419       & \textbf{95.2\%}         & 0.427      &          84.6\%                     &   0.511                         \\
LightRAG (LLaMA)                           & 69.3\%         & 0.359       & 93.5\%         & 0.361      &     86.2\%                          &       0.422                     \\
LiteSemRAG                                 & \textbf{73.5\%}         & \textbf{0.694}       & 92.3\%         & \textbf{0.692}      & \textbf{88.3\%}                        & \textbf{0.806}                      \\ \hline
\end{tabular}
\end{table}
Table~\ref{retrieval results} reports Recall@10 and MRR@10 on three datasets. 
Since GraphRAG does not provide explicit ranking of retrieved chunks, MRR@10 cannot be computed for GraphRAG. Therefore, we report only Recall@10 for GraphRAG. Additionally, when using LLaMA3-8B-Instruct as the underlying LLM, GraphRAG consistently encountered \textbf{Exceed maximum Token Limitation} (ETL) errors across all datasets, even though the chunk size was very small and contained only a few sentences (the maximum context length of LLaMA3-8B-Instruct is 8192 tokens). This error occurs not only during the indexing stage but also when querying databases that were already indexed using GPT. It prevented successful evaluation under this setting. Consequently, we mark these entries as ETL and exclude them from the comparison. This behavior reflects the heavy reliance of GraphRAG on LLM-based operations during indexing and querying. Experimental results show that LiteSemRAG achieves the best MRR@10 across all three datasets. In terms of Recall@10, LiteSemRAG achieves the top performance on \emph{HotpotQA} and \emph{FEVER}, and remains competitive on \emph{SciFact}. These results demonstrate that LiteSemRAG provides strong retrieval effectiveness while operating without LLM-dependent indexing and querying.

\subsection{Efficiency Analysis}
We report indexing and query efficiency on the three datasets in 
Table~\ref{tab:Time_results} and Table~\ref{tab:token_results}.  The experimental results show that LiteSemRAG substantially improves both indexing and query efficiency compared to LightRAG and GraphRAG. In particular, GraphRAG exhibits significantly higher query latency and token consumption, largely due to its LLM-dependent reasoning and graph traversal process. Across the evaluated datasets, LiteSemRAG achieves approximately 20$\times$ faster indexing speed and 10$\times$ faster querying speed compared to LightRAG. When compared with GraphRAG, LiteSemRAG achieves approximately 10$\times$ faster indexing speed and 100$\times$ faster querying speed. In terms of token consumption, both LLM-based RAG systems require approximately \emph{10K} tokens to index a single document on average, indicating substantial computational and monetary resource cost. In contrast, LiteSemRAG does not rely on LLMs during indexing or querying, eliminating LLM-related overhead entirely.

\begin{table}[t]
\centering
\caption{Indexing and query time efficiency comparison across datasets}
\label{tab:Time_results}

\begin{tabular}{lcccccc}
\hline
                 & \multicolumn{2}{c}{HotpotQA}    & \multicolumn{2}{c}{SciFact}     & \multicolumn{2}{c}{Fever}       \\ \cline{2-7} 
Method           & AIT            & AQT            & AIT            & AQT            & AIT            & AQT            \\ \hline
GraphRAG (GPT)   & 3.55s          & 29.8s          & 4.35s          & 48.76s         & 3.08s          & 30.66s         \\
GraphRAG (LLaMA) & ETL            & ETL            & ETL            & ETL            & ETL            & ETL            \\
LightRAG (GPT)   & 7.98s          & 0.58s          & 10.31s         & 1.57s          & 7.71s          & 1.45s          \\
LightRAG (LLaMA) & 8.06s          & 1.72s          & 10.39s         & 1.33s          & 8.12s          & 1.13s          \\
LiteSemRAG       & \textbf{0.24s} & \textbf{0.11s} & \textbf{0.69s} & \textbf{0.13s} & \textbf{0.47s} & \textbf{0.09s} \\ \hline
\end{tabular}
\end{table}

\begin{table}[t]
\centering
\caption{LLM token consumption comparison across datasets}
\label{tab:token_results}
\begin{tabular}{lcccccc}
\hline
                 & \multicolumn{2}{c}{HotpotQA} & \multicolumn{2}{c}{SciFact} & \multicolumn{2}{c}{Fever} \\ \cline{2-7} 
Method           & ITC           & QTC          & ITC          & QTC          & ITC         & QTC         \\ \hline
GraphRAG (GPT)   & 8.07K         & 10.85K       & 9.92K        & 11.24K       & 7.11K       & 4.06K       \\
GraphRAG (LLaMA) & ETL           & ETL          & ETL          & ETL          & ETL         & ETL         \\
LightRAG (GPT)   & 7.30K         & 0.43K        & 9.32K        & 0.59K        & 8.58K       & 0.44K       \\
LightRAG (LLaMA) & 8.17K         & 0.47K        & 8.85K        & 0.56K        & 8.23K       & 0.45K       \\
LiteSemRAG       & \textbf{0}    & \textbf{0}   & \textbf{0}   & \textbf{0}   & \textbf{0}  & \textbf{0}  \\ \hline
\end{tabular}

\end{table}

\subsection{Ablation Study}
To evaluate the contribution of the two-step semantic-aware retrieval strategy, we conduct an ablation study using a broad search and reranking retrieval variant. In this setting, we first retrieve all semantic nodes matched by the query. All chunks connected to these semantic nodes are then collected, and a \textbf{jina-reranker-v3} reranker is applied to rank the candidate chunks. If the number of candidate chunks becomes extremely large, we randomly sample 100 chunks as reranking candidates. The top-10 chunks are selected as the final results. This ablation removes the two-step semantic-aware retrieval mechanism while preserving the same semantic index graph and matching process. In particular, both methods use the identical index graph constructed during the indexing stage and rely on the same token-level text encoder to generate contextual embeddings. The only difference lies in the retrieval strategy applied at query time. Table~\ref{tab:ablation_results} compares the standard LiteSemRAG and the broad search and reranking variant in terms of Recall@10 (R@10), MRR@10 (M@10), and Average Query Time (AQT). Since both methods operate on the same index graph, the comparison isolates the effect of the retrieval strategy itself. The results show that the proposed two-step semantic-aware retrieval significantly outperforms the broad retrieval baseline, demonstrating that the performance gains of LiteSemRAG primarily arise from the semantic-aware co-occurrence graph retrieval mechanism rather than the use of the token-level text encoder alone.

\begin{table}[htb]
\centering
\caption{Ablation study on two-step semantic-aware retrieval}
\label{tab:ablation_results}
\begin{tabular}{cccccccccc}
\hline
\multirow{2}{*}{Method}  & \multicolumn{3}{c}{HotpotQA} & \multicolumn{3}{c}{SciFact} & \multicolumn{3}{c}{Fever} \\ \cline{2-10} 
                         & R@10     & M@10    & AQT     & R@10     & M@10    & AQT    & R@10    & M@10   & AQT    \\ \hline
Broad search and reranking & 32.5\%   & 0.197   & 2.85s   & 14.9\%   & 0.056   & 2.97s  & 46.7\%  & 0.251  & 3.24s  \\
Two-step semantic-aware  & 73.5\%   & 0.694   & 0.24s   & 92.3\%   & 0.692   & 0.69s  & 88.3\%  & 0.806  & 0.47s  \\ \hline
\end{tabular}
\end{table}

\section{Conclusion}
In this paper, we proposed LiteSemRAG, a lightweight and fully LLM-free semantic-aware graph retrieval framework for Retrieval-Augmented Generation. LiteSemRAG constructs a heterogeneous semantic graph from contextual embeddings, explicitly separating surface lexical forms from context-dependent semantic meanings to model token polysemy without LLM-based indexing or reasoning. The framework introduces dynamic semantic node construction with chunk-level context aggregation and adaptive anomaly handling, together with a two-step semantic-aware retrieval strategy. Experiments on three benchmark datasets show that LiteSemRAG achieves competitive or superior retrieval performance compared to LLM-based graph RAG systems while substantially improving indexing and querying efficiency and eliminating LLM token consumption.

%
%
%
\bibliographystyle{splncs04}
\bibliography{references}

@article{lewis2020retrieval,
  title={Retrieval-augmented generation for knowledge-intensive nlp tasks},
  author={Lewis, Patrick and Perez, Ethan and Piktus, Aleksandra and Petroni, Fabio and Karpukhin, Vladimir and Goyal, Naman and K{\"u}ttler, Heinrich and Lewis, Mike and Yih, Wen-tau and Rockt{\"a}schel, Tim and others},
  journal={Advances in neural information processing systems},
  volume={33},
  pages={9459--9474},
  year={2020}
}

@inproceedings{karpukhin2020dense,
  title={Dense passage retrieval for open-domain question answering},
  author={Karpukhin, Vladimir and Oguz, Barlas and Min, Sewon and Lewis, Patrick and Wu, Ledell and Edunov, Sergey and Chen, Danqi and Yih, Wen-tau},
  booktitle={Proceedings of the 2020 conference on empirical methods in natural language processing (EMNLP)},
  pages={6769--6781},
  year={2020}
}

@inproceedings{sarthi2024raptor,
  title={Raptor: Recursive abstractive processing for tree-organized retrieval},
  author={Sarthi, Parth and Abdullah, Salman and Tuli, Aditi and Khanna, Shubh and Goldie, Anna and Manning, Christopher D},
  booktitle={The Twelfth International Conference on Learning Representations},
  year={2024}
}

@article{zhao2026retrieval,
  title={Retrieval-augmented generation for ai-generated content: A survey},
  author={Zhao, Penghao and Zhang, Hailin and Yu, Qinhan and Wang, Zhengren and Geng, Yunteng and Fu, Fangcheng and Yang, Ling and Zhang, Wentao and Jiang, Jie and Cui, Bin},
  journal={Data Science and Engineering},
  pages={1--29},
  year={2026},
  publisher={Springer}
}

@inproceedings{devlin2019bert,
  title={Bert: Pre-training of deep bidirectional transformers for language understanding},
  author={Devlin, Jacob and Chang, Ming-Wei and Lee, Kenton and Toutanova, Kristina},
  booktitle={Proceedings of the 2019 conference of the North American chapter of the association for computational linguistics: human language technologies, volume 1 (long and short papers)},
  pages={4171--4186},
  year={2019}
}

@article{brown2020language,
  title={Language models are few-shot learners},
  author={Brown, Tom and Mann, Benjamin and Ryder, Nick and Subbiah, Melanie and Kaplan, Jared D and Dhariwal, Prafulla and Neelakantan, Arvind and Shyam, Pranav and Sastry, Girish and Askell, Amanda and others},
  journal={Advances in neural information processing systems},
  volume={33},
  pages={1877--1901},
  year={2020}
}

@inproceedings{borgeaud2022improving,
  title={Improving language models by retrieving from trillions of tokens},
  author={Borgeaud, Sebastian and Mensch, Arthur and Hoffmann, Jordan and Cai, Trevor and Rutherford, Eliza and Millican, Katie and Van Den Driessche, George Bm and Lespiau, Jean-Baptiste and Damoc, Bogdan and Clark, Aidan and others},
  booktitle={International conference on machine learning},
  pages={2206--2240},
  year={2022},
  organization={PMLR}
}

@inproceedings{Wadden2020FactOF,
  title={Fact or Fiction: Verifying Scientific Claims},
  author={David Wadden and Shanchuan Lin and Kyle Lo and Lucy Lu Wang and Madeleine van Zuylen and Arman Cohan and Hannaneh Hajishirzi},
  booktitle={EMNLP},
  year={2020},
}

@inproceedings{yang2018hotpotqa,
  title={HotpotQA: A dataset for diverse, explainable multi-hop question answering},
  author={Yang, Zhilin and Qi, Peng and Zhang, Saizheng and Bengio, Yoshua and Cohen, William and Salakhutdinov, Ruslan and Manning, Christopher D},
  booktitle={Proceedings of the 2018 conference on empirical methods in natural language processing},
  pages={2369--2380},
  year={2018}
}

@article{thakur2021beir,
  title={Beir: A heterogenous benchmark for zero-shot evaluation of information retrieval models},
  author={Thakur, Nandan and Reimers, Nils and R{\"u}ckl{\'e}, Andreas and Srivastava, Abhishek and Gurevych, Iryna},
  journal={arXiv preprint arXiv:2104.08663},
  year={2021}
}

@inproceedings{izacard2021leveraging,
  title={Leveraging passage retrieval with generative models for open domain question answering},
  author={Izacard, Gautier and Grave, Edouard},
  booktitle={Proceedings of the 16th conference of the european chapter of the association for computational linguistics: main volume},
  pages={874--880},
  year={2021}
}

@inproceedings{guu2020retrieval,
  title={Retrieval augmented language model pre-training},
  author={Guu, Kelvin and Lee, Kenton and Tung, Zora and Pasupat, Panupong and Chang, Mingwei},
  booktitle={International conference on machine learning},
  pages={3929--3938},
  year={2020},
  organization={PMLR}
}

@inproceedings{khattab2020colbert,
  title={Colbert: Efficient and effective passage search via contextualized late interaction over bert},
  author={Khattab, Omar and Zaharia, Matei},
  booktitle={Proceedings of the 43rd International ACM SIGIR conference on research and development in Information Retrieval},
  pages={39--48},
  year={2020}
}

@article{yunollmrag,
  title={NoLLMRAG: LLM-Free Makes Graph-Based RAG Highly Efficient, Effective and Generalizable},
  author={Yu, ChengZhuo and Su, Fang}
}

@article{guo2024lightrag,
  title={Lightrag: Simple and fast retrieval-augmented generation},
  author={Guo, Zirui and Xia, Lianghao and Yu, Yanhua and Ao, Tian and Huang, Chao},
  journal={arXiv preprint arXiv:2410.05779},
  volume={2},
  number={3},
  year={2024}
}

@article{gutierrez2024hipporag,
  title={Hipporag: Neurobiologically inspired long-term memory for large language models},
  author={Guti{\'e}rrez, Bernal J and Shu, Yiheng and Gu, Yu and Yasunaga, Michihiro and Su, Yu},
  journal={Advances in neural information processing systems},
  volume={37},
  pages={59532--59569},
  year={2024}
}

@article{edge2024local,
  title={From local to global: A graph rag approach to query-focused summarization},
  author={Edge, Darren and Trinh, Ha and Cheng, Newman and Bradley, Joshua and Chao, Alex and Mody, Apurva and Truitt, Steven and Metropolitansky, Dasha and Ness, Robert Osazuwa and Larson, Jonathan},
  journal={arXiv preprint arXiv:2404.16130},
  year={2024}
}

\end{document}